\RequirePackage{fix-cm}
\documentclass[smallextended]{svjour3}       
\smartqed  

\usepackage{hyperref}
\usepackage{amssymb}
\usepackage{graphicx}
\usepackage{amsmath}
\usepackage{multirow}
\usepackage{mathtools}
\usepackage[english]{babel}
\usepackage{amsmath}
\usepackage{rotating}
\usepackage{array}
\usepackage{float}
\usepackage{subcaption}

\usepackage[utf8]{inputenc}
\usepackage[english]{babel}

 \usepackage{graphicx,fancybox}
\usepackage{mathtools}

 \usepackage{framed}

\usepackage{caption}

\usepackage[numbers,compress]{natbib}
\usepackage[a4paper, margin=3cm]{geometry}

\begin{document}

\title{SPSMAT: GNU Octave software package for spectral and pseudospectral methods}

\titlerunning{ SPSMAT software package}        

\author{Sobhan Latifi$^1$ and Mehdi Delkhosh$^2$}

\institute{
$^1$ Department of Computer Sciences, Shahid Beheshti University, Tehran, Iran. \email{s.latifi@mail.sbu.ac.ir}.\\            
$^2$Department of Mathematics and Computer Science, Islamic Azad University, Bardaskan Branch, Bardaskan, Iran.
\email{mehdidelkhosh@yahoo.com}.\\
}
\date{Received: date / Accepted: date}
\maketitle
\begin{abstract}
SPSMAT (Spectral/Pseudospectral matrix method) is an add-on for Octave, that helps you solve nonfractional-/fractional ordinary/partial  differential/integral equations. In this version, as the first version, the well-defined spectral or pseudospectral algorithms are considered to solve differential and integral equations. The motivation is that there are few software packages available that make such methods easy to use for practitioners in the field of scientific computing. Additionally, one of the most practical platforms in computation, MATLAB,
is currently not supporting beneficial and free numerical method for the solution of differential equations--to the best  author's knowledge. To remedy this situation, this paper provides the description of its relevant uploaded open source software  package and is a broad guidance to  describe how to work with this toolbox.  
\end{abstract}
\section{Introduction} 

During the recent years getting our hands dirty of working with spectral and pseudospectral methods, when it comes to solving  differential/integral equations, we feel like there is a huge lack amongst a myriad of papers published in the field of solving differential/integral equations. The instant problem to detect is that all the scientists and researchers solve problems without using others' codes and packages. Undoubtedly, having a fixed platform and inviting the scientist to develop their work  can reduce the time and effort in solving these similar-based problems. 
Over these years, we have known that the works of these scientists are of giant works and worth admiring, but we always ask ourself "what if we gather the works of all current scientific computer scientist  in order  to design a package to use easier and grownup methods for the solutions of ODEs and PDEs?".  
\\
To do so, it has been attempted to provide  consistent and user-friendly high-level functions that allow experimental scientists to work properly and beneficially on their equations.\\
Furthermore, users and developers can easily extend the functionality and implement new
algorithms due to the modular design functionalities and facilities, and in higher expectation,  harness them in other software packages.

This package and its documentation are freely
available from \href{https://github.com/siaadfilml/SPSMAT}{This address}.

The SPSMAT toolbox consists of approximately 80
high-level and other low-level functions.
We are to say that the high-level functions give a consistent
and easy-to-use interface of the functionality and reusability to the users and developers, allowing them to do the analysis in such well-defined manners. The
low-level functions implement the core functionality and are not needed to be shown to the end-users, these functions are set and called inside  other functions. In addition to handy functions, some constructive examples are given in terms of guidance and showing the  handling of the functions.
\\SPSMAT has been believed to be portable and open source. At this
aim, SPSMAT has been developed using Octave, which is a kind of  free MATLAB clone and runs on the
commonest operating systems, such as Unix, Linux, Windows, and Mac OS X. \\
To the best of our knowledge, SPSMAT is also the first free scientific computing package which runs on GNU/Octave platform. 

\section{Features of SPSMAT}

The SPSMAT package is
self-contained; it only needs an Octave environment
with
standard toolboxes [See
 \href{http://www.gnu.org/software/octave/} {Octave} ]. 
SPSMAT is released under the free and open source       GNU GENERAL PUBLIC LICENSE (GPL)
(Version 3, 29 June 2007).  The accompanying source code and the documentation of the toolbox have been enriched with numerous comments, examples, and detailed instructions for extensions.
\\
This is the first in a series of papers on implementing different numerical techniques in term of  an open source
GNU Octave toolbox. The intention of this ongoing project is
to provide a rapid prototyping package
for  using spectral and pseudospectral methods in solving integral and differential equations. 
\\
Here, we highlighted some of the outstanding features of SPSMAT:
\begin{itemize}
\item  highly modular; 
the reasons beyond  highly modular design are:
\begin{itemize}
\item * The combinations of the approximation techniques.
\item  * The easy construction and embedding of other novel techniques; the library aims to utilize the well-known spectral and pesudospectral methods as transparent as possible to the user. A small alternation and enhancing the code to fit one's assumption is easy to handle.
\item * Having a well-defined structure for the modules also simplifies the
maintenance of the code.
 \item * Their immediate application in other 
  packages. The modular design facilitates reuse of source code in
other software. It ignites  the
collaboration with  current developers and with developers
of other software packages to broaden their work and understandings.
\end{itemize}

\item  Free and open source; explained in next sections.
\item  Multi-platform toolbox; SPSMAT is
multi-platform: it has been extensively tested on Windows and Linux; since it is made of standard Octave  files, it is expected to work on
alternative platforms as well.
\item Following "all-in-one"  philosophy. By this, we tried to represent one particular function  with different inputs to encompass all possible outputs and purposes.
For instance, in order to have functions for Jacobi polynomials and their derivatives, we defined a function with input 0 to show the Jacobi polynomials and with input 1 to give the first order derivative of this polynomials.

\end{itemize}

\subsection{Why numerical methods?}

After delving through different books, we know that
they concentrated on ODEs based on
linear problems. 
  The trouble is most ODE problems including almost
all nonlinear ones can't be solved analytically.  
A numerical one, however, can be obtained by  computer. The very important advantage of numerical approximations is that they can be obtained for almost any ODE. In addition to the speed of these methods, the nature of numbers helps scrutinize results
obtained numerically  even  apply
different  analysis method to a problem and compare them with one another. On another hand, it is not easy  to compare two exact formulas, but it is beneficial to compare two numbers, even  as numbers
or visually in  plots \cite{therifeten}. That simply can be detected as a reason for designing a numerical package rather than the analytical one. 

\subsection{Why No User Interface (UI)?}
We never assume that we are familiar with everything that a user
might want to do.
A great feature of the
SPSMAT toolbox is that it does not have a Graphical User
Interface (GUI). Instead, the user is interacting directly with
the functions on the  command line or in scripts and is able to mix and build his new insight in order to solve a problem. 
Providing a GUI prevent the user to 
mix, match, and test the SPSMAT high-level functions experimentally. 
Data and code, in fact, are in the hands of the end users to reshape them suitable to their need. 
\subsection{Why Open source and Octave?}
MATLAB is widely
known and used in the science communities including scientific computing. Although
MATLAB has a rich feature set and flexibility, it is  relatively and ridiculously   expensive. MATLAB environment is a commercial and "closed" product, thus MATLAB
kernel and libraries cannot be modified nor freely distributed.
To allow exchanging insights and contributing scientific
research, both the toolbox and the platform on which a toolbox runs must be completely free.

Mathworks like other proprietary firms has its own policies that may not help you freely use and develop some packages.

 
That was maybe the ultimate motive to use an open source platform. In addition to it, as what an Iranian proverb saying \textbf{ one clap is not audible}, we
 required
the users extend the analyses with their own code and be exchanged between users, and between students and
their supervisors, facilitating collaboration and knowledge over Github where they can even transfer their proper contribution with us. \\Nonetheless,  these are not such limitation for MATLAB users at all. They can even download the package and use it.

\subsection{Why Fractional?}
Due to the accuracy of fractional differential and integral equations in modeling various natural phenomena, fractional calculus has become the focus of many
researchers. 
Fractional differential equations provide
an outstanding instrument to describe the complex phenomena in fields of viscoelasticity, electromagnetic waves, diffusion equations, and so on. Moreover, the fractional
order models of real systems are more sufficient in comparison with the integer order
cases. Therefore, the field of fractional calculus has motivated the interest of researchers
in various fields like physics, chemistry, engineering and even finance where all of their scientists and students need to work with  numerical methods of solving these fractional equations. We broadly aim at this spectrum and wanted to produce a package for all of these practitioners. 
\subsection{Why Jacobi?}
The classical Jacobi polynomials, have been used extensively in mathematical
analysis and practical applications, and play a pivotal role in the analysis and implementation of spectral and pseudospectral methods \cite{bhrawy2}.
It is easily proven that the Jacobi polynomials are  the  polynomials arising as eigenfunctions of a singular
Sturm--Liouville problem  \cite{bhrawy2}. 
The usual spectral method by Legendre or Chebyshev approximation, and available only for non-singular problems on rectangular bounded domains. However, the general Jacobi method can be used in a wide array of
problems \cite{bhrawy3}.\\
In addition to this, in this version, also for the simplicity's sake, we cover Jacobi polynomials which are  wide-ranging polynomials encompassing Gegenbauer (ultraspherical), Chebyshev, and Legendre polynomials.  But, for those who are not into these polynomials, we also offered the derived subcategories  including Gegenbauer, Chebyshev (all four kinds), and Legendre polynomials. 
\subsection{Why Now?}
Over the recent years, many researchers have been interested in studying the properties
of various equations and  providing robust and accurate analytical and numerical
methods for solving these equations and recorded their work in their published papers: what we have seen at this moment is to see and gather their works in one single work.
Moreover, some of the high-qualified state-of-the-art methods and landmark works are done and  analyzed perfectly that let us code them in the best shape that is readable and understandable, and specifically, beneficial.  In addition to this, other works done on spectral methods are out-dated; like the work of Funaro/Matlab  that is done in 1993 via Fortran (See \href{http://cdm.unimo.it/home/matematica/funaro.daniele/rout.htm} {Funaro package}). What these packages lack are that at that time fractional calculus, operational matrices and Lagrangian bases had not been developed like today. The other FORTRAN package, PseudoPack 2000 is implemented by Don and Costas
and is available in
[\href{http://www.cfm.brown.edu/people/wsdon/home.html}{PseudoPack package}]. In this work parallel computers capability involved to solve problems and suffer the same misfortunes we issued here. In the newest work by Trefethen in 
\href{http://www.chebfun.org/docs/guide/}{Chebfun package},
a huge and broad work done to refer a tool for researchers. But the light version, and more importantly, the specification for solving differential and integral equation in terms of spectral and pseudospectral was not its first goal. 

SPSMAT, lo and behold, was created to address some of these issues.

\subsection{Why Matrix method?}
As the operations in MATLAB and Octave are based on the concept of matrix, and in some extends their performance is optimal for matrices, we have given a 
matirx-based method in order to reach the best action of MATLAB or Octave.
Another reason is the parallel computation that is said to be feasible and meaningful when it comes to matrix operations like multiplications and so forth. In this version, we have not considered matrix parallel computation methods, but in due course, we will work on it and report its relevant results. 
\subsection{Why Spectral?}
The well-defined Spectral methods, by using computer power,  have been developed rapidly through the recent years
for the numerical solutions of ordinary/partial differential/integral equations. They have been providing a satisfying accuracy compared with
other numerical methods and have wide applications in many
mathematical problems. As we have realized, the  spectral methods have been developed in different ways to provide accurate solutions for linear and nonlinear ordinary/partial differential/integral equations. Additionally, spectral methods
have  received considerable attention in dealing with various problems. As their most significant characteristics,
they reduce  problems to those of solving a system of algebraic equations; this makes them easier for solving. Also,
they have excellent error properties and they offer exponential rates of convergence for smooth problems.
Such methods
have matured over the last 50 years and, in many cases, meet the robustness
and  computational  efficiency  standards  required  in  practice. 
\subsection{Why Pseudospectral?}
In general, there are two
ways to construct a polynomial approximation to the solution $y(x)$:
One is to use an interpolating polynomial between the values $y(x_j)$ at
some points $x_j$. The other is to use a series expansion in terms of  Lagrange  or orthogonal polynomials like Jacobi polynomials. Here, the latter idea is pursued,
and both the Jacobi and Lagrange Jacobi polynomials are constructed and the difference between employing them  are also represented. 
In fact, the pseudospectral methods (here using Lagrange polynomials) are based on the spectral techniques. In pseudospectral methods, there are particularly two steps to achieve a numerical approximation
for a  differential equation. Firstly, a proper finite or discrete representation of the
solution should be selected-- this can be done by polynomial interpolation of the solution based on
some appropriate nodes. On another hand, it is  known that the Lagrange interpolation polynomial based
on equidistant  points does not yield a satisfactory approximation to general smooth functions. As a matter of fact, when the number of collocation points increases, interpolant polynomials notably diverge.  However, satisfying  results are obtained
by relating the collocation points to the structure of classical orthogonal polynomials, such as the well-established Jacobi-Gauss-Lobatto points. That can be figured as a cogent reason for choosing these points in defining Lagrange polynomials. The use of global polynomials together with
Gaussian quadrature collocation points (here Jacobi-Gauss-Lobatto points) is known to provide accurate approximations that converge exponentially for problems
whose solutions are smooth \cite{kidderlatifi,fokkerlatifi,delkhosh,delkhosh2,div}.

\subsection{What is the prerequisite and who is this package for?}
This software package is intended for those who have a taste for solving   ODEs, PDEs, integral equations and Optimal Controls.  Practitioners with a not-so-deep understanding  of Matlab or Octave can use this package without any difficulty. But basically this package is meant to be useful to solve different differential and integral equations; so it stipulates that the users  have a grasp of Numerical analysis beforehand.

Undergraduate students taking a course in scientific computing and any relevant course can find this package as a test machine to check different things with orthogonal polynomials and develop their work with this package: this is your lightweight companion. In nutshell, whoever you are, we aim to increase your appreciation of this fundamental subject.
\section{Solving a  nonlinear PDE with SPSMAT}
In this section, firstly we discuss the orthogonal Gegenbauer polynomials. The orthogonal Gegenbauer polynomial of order  $N$ is defined over $[-1,1]$  \cite{bahram}:

\begin{equation}
G_{N}^{\alpha}(x)=\sum_{i=0}^{\lfloor\frac{N}{2}\rfloor}(-1)^i\frac{\Gamma(N+\alpha-i)}{\Gamma(i!(N-2i)!\Gamma(\alpha))}(2x)^{N-2i},
\end{equation}
\begin{equation}
~G_{N}^{\alpha}(-x)=(-1)^NG_{N}^{\alpha}(x),~G_{N}^{\alpha}(\pm 1 )=(\pm 1)^N\frac{\Gamma(N+2\alpha)}{\Gamma(2\alpha)\Gamma(N+1)}.
\end{equation}
Other features of Gegenbauer polynomials can be found in \cite{bahram}.\\

As said, this package helps solve partial and ordinary differential equations. To show its capability,  we have considered nonlinear Fokker-Plank equation:
$$ \frac{\partial y}{\partial t} =\big[\frac{-\partial }{\partial x}A(x,t,y)+ \frac{\partial^2 }{\partial x^2}B(x,t,y)\big]y,$$
 where $ A(x,t,y)= \frac{4y}{x}-\frac{x}{3}$ and $ B(x,y,t)=y $ with the initial  and boundary conditions $y(x,0)=x^2$, $y(0,t)=0$ and $y(1,t)=e^t$. The exact solution is $y(x,t)=x^2e^t$ and $x\in [0,1],~t\in [0,1]$. 

For solving this problem, with help of Crank-Nicolson method \cite{fokkerlatifi}, we discretized the variable $t$ and solved the equation in different iterations \cite{fokkerlatifi}.
Notice that $t_n=n\times\bigtriangleup t$ and $\bigtriangleup t=t_{n+1}-t_n$.\\
Using $\frac{\partial y}{\partial t}=\frac{y^{n+1}-y^{n}}{\bigtriangleup t}$ where $y^{n}=y(x,t_n)$ and applying Crank-Nicolson method we have got
$$ y_t=(\frac{4y}{x^2}-\frac{8y_x}{x}+\frac{1}{3}+2y_{xx})y+(\frac{x}{3}+2y_x)y_x, $$
\begin{eqnarray}
&&\frac{y^{n+1}-y^{n}}{\bigtriangleup t}=\theta (\frac{4y^{n}}{x^2}-\frac{8y_x^{n}}{x}+\frac{1}{3}+2y_{xx}^{n})y^{n+1}\nonumber\\&&+(\frac{x}{3}+2y_x^{n})y_x^{n+1}+\bigtriangleup t(1-\theta) (\frac{4y^{n}}{x^2}-\frac{8y_x^{n}}{x}+\frac{1}{3}+2y_{xx}^{n})y^{n}+(\frac{x}{3}+2y_x^{n})y_x^{n},
\end{eqnarray}
in which $y^{n+1}$ is unknown and is supposed to be determined. Simplifying the last equation we have
\begin{eqnarray}\label{eqq}
&&\bigg[1-\theta\bigtriangleup t\bigg((\frac{4y^{n}}{x^2}-\frac{8y_x^{n}}{x}+\frac{1}{3}+2y_{xx}^{n})\bigg)\bigg]y^{n+1}-\theta\bigtriangleup t(\frac{x}{3}+2y_x^{n})y_{x}^{n+1}=y^{n}+\nonumber\\&&
(1-\theta)\bigtriangleup t \bigg((\frac{4y^{n}}{x^2}-\frac{8y_x^{n}}{x}+\frac{1}{3}+2y_{xx}^{n})\bigg)y^{n} +(1-\theta)\bigtriangleup t (\frac{x}{3}+2y_x^{n})y_{x}^{n}.
\end{eqnarray}
 Now by defining
$$s_1:=x\rightarrow \bigtriangleup t (\frac{4y^{n}}{x^2}-\frac{8y_x^{n}}{x}+\frac{1}{3}+2y_{xx}^{n}),$$
$$s_2:=x\rightarrow \bigtriangleup t (\frac{x}{3}+2y_x^{n}),$$
then Eq. (\ref{eqq}) will be 
\begin{eqnarray}\label{eqq_}
&&\bigg(1-\theta s_1(x)\bigg)y^{n+1}-\theta s_2(x) y_{x}^{n+1}=\bigg(1+
(1-\theta)s_1(x)\bigg)y^{n} +(1-\theta)s_2(x)y_{x}^{n}.\nonumber
\end{eqnarray}
Now for solving this equation, we use an approximation expansion with Generalized Lagrange polynomials:
$y^{n+1}=\sum_{i=0}^{N}L_{i}^{u}(x)C_i^{n+1}$, where $u=2x-1$,  $C_i^{n+1}$ are the unknown coefficients, and $L_{i}^{u}(x)$ are the Generalized Lagrange polynomials defined in \cite{kidderlatifi,fokkerlatifi}. Then

\begin{eqnarray}\label{eqq__}
&&\bigg(1-\theta s_1(x)\bigg)\sum_{i=0}^{N}L_{i}^{u}(x)C_i^{n+1}-\theta s_2(x) \sum_{i=0}^{N}\big(L_{i}^{u}(x)\big)'C_i^{n+1}=\nonumber\\&&
\bigg(1+
(1-\theta)s_1(x)\bigg)\sum_{i=0}^{N}L_{i}^{u}(x)C_i^{n} +(1-\theta)s_2(x)\sum_{i=0}^{N}\big(L_{i}^{u}(x)\big)'C_i^{n},
\end{eqnarray}

and with Gegenbauer Gauss Lobatto points $x_0=0$, $x_{N}=1$ and $N-1$ roots of $G_{N-1}^{\theta}(u)$ which are calculated with:
$$x_i=[0,roots(G_{N-1}^{\alpha+1}(u)),1],i=0..N.$$
\begin{center}
\begin{figure}[H]
\centering
  \fbox{
\includegraphics[scale=0.46]{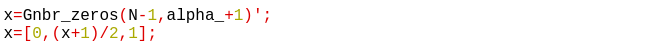}}
\caption{Gegenbauer Gauss Lobatto points}
\label{piece2}
\end{figure}
\end{center}
Setting 
$$v_1=[s_1(x_1),...,s_1(x_{N-1})]^T,$$
$$v_2=[s_2(x_1),...,s_2(x_{N-1})]^T,$$
$$m_1=[1,1,\dots,1]^T-\theta v_1, m_2=\theta v_2,$$
$$\hat{m_1}=[1,1,\dots,1]^T-(1-\theta) v_1, \hat{m_2}=(1-\theta) v_2,$$
 and considering Eq. (\ref{eqq__}), by using  Gegenbauer Gauss Lobatto points, we will get a matrix form as 
 \begin{equation}\label{eqqq}
 A^{n+1}C^{n+1}=b^{n}
 \end{equation}
 in which $A^{n+1}$ is the coefficient matrix and $ C^{n+1} $ is the unknown vector.

By implementing boundary condition, $ A^{n+1} $ is rewritten as: 
\[
A^{n+1}=\begin{bmatrix}
A_1\\diag(m_1)\times A_2-diag(m_2)\times A_5\\
A_3
\end{bmatrix},
b^{n}=\begin{bmatrix}
0\\diag(\hat{m_1})\times A_2+diag(\hat{m_2})\times A_5\\
e^t
\end{bmatrix}
\]
 by which 
\[
A_1=\begin{bmatrix}
1&0&\dots&0
\end{bmatrix},
A_3=\begin{bmatrix}
0&0&\dots&1
\end{bmatrix},
A_2=\begin{bmatrix}
0&1&0&\dots&0\\
0&0&1&\dots&0\\
\vdots\\
0&0&\dots&1&0\\
\end{bmatrix},
\]
\[
A_4=D^{(2)}[1:N-1,:], ~A_5=D^{(1)}[1:N-1,:],
\]
 
where $ D^{(1)} $ and $D^{(2)}$ are the Generalized Lagrange first and second order derivative matrices that are calculated in \cite{kidderlatifi,fokkerlatifi}.
 
\begin{center}
\begin{figure}[H]
\centering
  \fbox{
\includegraphics[scale=0.46]{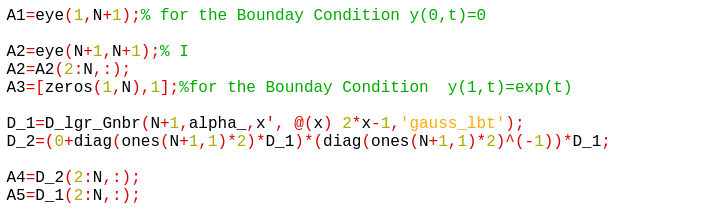}}
\caption{Piece of code for obtaining $A_1\dots A_5$}
\label{piece2}
\end{figure}
\end{center}
 
\begin{center}
\begin{figure}[H]
\centering
  \fbox{
\includegraphics[scale=0.46]{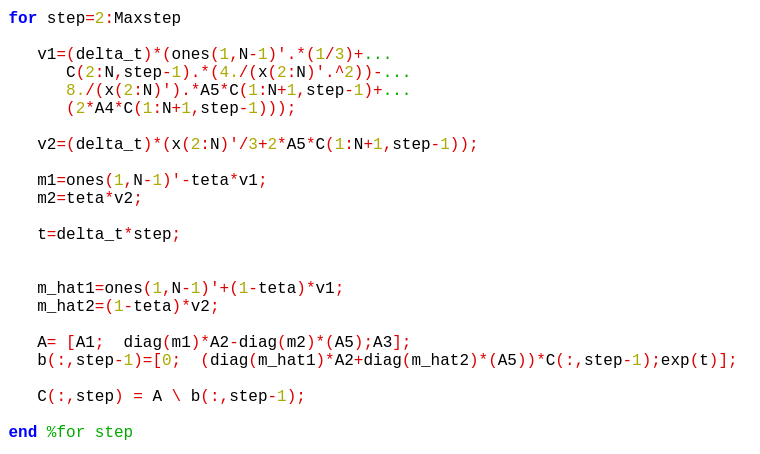}}
\caption{Solving $AC^{n+1}=b^n$ at different time steps.}
\label{piece2}
\end{figure}
\end{center}
  
 Also  applying initial condition $y^0=x^2$ leads to $C^0=[x_0^2,x_1^2...,x_N^2]$ .
 \begin{center}
\begin{figure}[H]
\centering
  \fbox{
\includegraphics[scale=0.46]{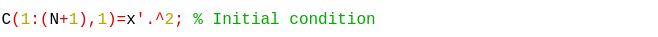}}
\caption{Initial condition}
\label{piece2}
\end{figure}
\end{center}
 
\begin{center}
\begin{figure}
\centering  
 \includegraphics[scale=0.48]{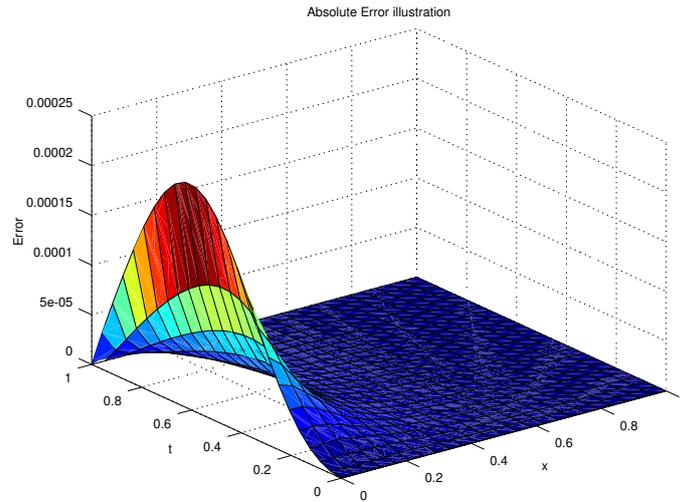}
 \caption{Error illustration.$\theta=0.5$,$\alpha=0.5$, $N=7$. }
 \label{exa6_error}
\end{figure}
\end{center}

 The result of solving the system  in Eq. (\ref{eqqq}) with SPSMAT is depicted in Fig. \ref{exa6_error}. 

 The completed code file contains more of these functions which are mostly discussed  \href{https://github.com/siaadfilml/SPSMAT}{here}.


\begin{acknowledgements}
In the end, we owe  some researchers a thank who were at our disposal when we were in dire need. We thank profusely  Mohammad Hemami and Dr. Saeed Kazem. 
\end{acknowledgements}

\bibliographystyle{spmpsci}      


\end{document}